\documentclass[a4paper,11pt]{article}
\usepackage{pos}
\usepackage{slashed}
\usepackage{caption}
\usepackage{subcaption}

\title{Phenomenology of electroweak spin-1 resonances}
\author[a]{R.~Caliri}
\emailAdd{rosy.caliri@uni-wuerzburg.de}
\author[a]{J.~Hadlik}
\emailAdd{jan.hadlik@uni-wuerzburg.de}
\author[a]{M.~Kunkel}
\emailAdd{manuel.kunkel@uni-wuerzburg.de}
\author*[a]{W.~Porod}
\emailAdd{werner.porod@uni-wuerzburg.de}
\author[b]{Ch.~Verollet}
\emailAdd{c.verollet@ip2i.in2p3.fr}

\affiliation[a]{Institut f\"{u}r Theoretische Physik und Astrophysik, University of  W\"{u}rzburg,\\ Emil-Hilb-Weg 22, D-97074 W\"{u}rzburg, Germany}

\affiliation[b]{Institut de Physique des 2 Infinis de Lyon (IP2I),\\ 69100 Villeurbanne Cedex,
France}

\abstract{Composite Higgs models with  a fermionic UV completion predict the existence
of various bound states. 
We investigate  models containing  SU(2)$_L\times$SU(2)$_R$ as part of the unbroken global subgroup
in the new strong sector. These models predict that there
are two neutral and one charged  spin-1 resonances mixing seizable with the SM vector bosons. 
These can be singly produced at the LHC. We explore
their LHC phenomenology and demonstrate that there are
still viable scenarios consistent with existing LHC data where the masses of these states 
can be as low as about 1.5 TeV. }

\FullConference{Proceedings of the Corfu Summer Institute 2025 "School and Workshops on Elementary Particle Physics and Gravity" (CORFU2025)\\
27 April - 28 September, 2025\\
Corfu, Greece\\}

\newcommand{\gt}{\Tilde{g}}
\newcommand{\gV}{g_{V\pi\pi}}
\newcommand{\hc}{\mathrm{h.c.}} 
\newcommand{\SO}{\mathrm{SO}}
\newcommand{\SU}{\mathrm{SU}}

\newcommand{\Sp}{\mathrm{Sp}}

\def\lsim{\mathrel{\rlap{\lower4pt\hbox{\hskip1pt$\sim$}}
    \raise1pt\hbox{$<$}}}         
\def\gsim{\mathrel{\rlap{\lower4pt\hbox{\hskip1pt$\sim$}}
    \raise1pt\hbox{$>$}}}

\begin{document}
\maketitle

\section{Introduction}

In composite Higgs models  \cite{Kaplan:1983fs,Kaplan:1983sm} one postulates an additional gauge interaction besides the ones of the Standard Model (SM). These models provide a potential 
solution to the problem of hierarchy between the electroweak (EW) scale and the Planck scale provided the new interaction become strong in the multi-TeV range leading to the spontaneous breaking of the global symmetries of the new fermions charged under this gauge interaction. The corresponding breaking scale is thus dynamically generated like in quantum chromodynamics (QCD). This breaking should be such, that the Higgs boson emerges as 
a pseudo Nambu Goldstone boson (pNGBs) \cite{Contino:2003ve} and its potential is generated by explicit
breaking terms: the gauging of the electroweak symmetry, the couplings of the top quark
\cite{Agashe:2004rs} and a 
possible mass term for the underlying fermions \cite{Galloway:2010bp,Cacciapaglia:2014uja}.

The size of the top-mass can be explained via partial
compositeness  \cite{Kaplan:1991dc}  leading to the prediction of top-partners, some of which will mix with
the SM quarks of the third generation. In addition, also spin-1 states are predicted, both in the coloured
 and the electroweak sector.
In this contribution, we focus on models based on an underlying 
gauge-fermion description as they allow for a systematic classification of the properties
of the resonances, see  ref.~\cite{Ferretti:2013kya} for a systematic way to obtain models
with  the minimal resonances required by the observed Higgs sector. 
An important class among these models contain two separate species in different irreducible representations
(irreps). A set of twelve minimal models, dubbed M1-M12, has been presented in 
ref.~\cite{Ferretti:2016upr,Belyaev:2016ftv}. 

The proposed models have served as a basis for several phenomenological studies:
for electroweak pNGBs see \cite{Ferretti:2016upr,Agugliaro:2018vsu,Cacciapaglia:2022bax,Flacke:2023eil},
for QCD coloured pNGBs \cite{Cacciapaglia:2015eqa,Belyaev:2016ftv,Cacciapaglia:2020vyf,Flacke:2025xwl}. Top 
partners with non-standard decays are discussed in \cite{Bizot:2018tds,Xie:2019gya,Cacciapaglia:2019zmj,Banerjee:2022izw,Banerjee:2024zvg} and those containing also color octets and sextets in \cite{Cacciapaglia:2021uqh,Cacciapaglia:2026jlv}.
Spin-1 resonances carrying electroweak 
charges are discussed in \cite{BuarqueFranzosi:2016ooy,Caliri:2024jdk} and those with QCD charges in \cite{Cacciapaglia:2024wdn}.
We note for completeness, that spectra and couplings for such resonances have been
computed on the lattice for models based on SU(4) 
\cite{Ayyar:2017qdf,Ayyar:2018glg,Ayyar:2018ppa,Ayyar:2018zuk,Ayyar:2019exp,Hasenfratz:2023sqa}, 
e.g.~models M6 and M11, and models based on Sp(4) 
\cite{Bennett:2017kga,Bennett:2019cxd,Bennett:2019jzz,Bennett:2022yfa,Kulkarni:2022bvh,Bennett:2023mhh,Bennett:2023qwx,Bennett:2024tex}, 
e.g.~models M5 and M8. Moreover, computations based on holography are also available 
\cite{Erdmenger:2020lvq,Erdmenger:2020flu,Elander:2020nyd,Elander:2021bmt,Erdmenger:2023hkl,Erdmenger:2024dxf}. 
Both approaches give consistent results for the covered subset of models.

In this contribution we focus on the phenomenology of electroweak charged spin-1 resonances, in particular
on those which mix with the SM vector bosons. In the subsequent sections we first summarize
 relevant model aspects and then focus on LHC phenomenology of the spin-1 resonances. We then discuss to which
extent this sector is constrained by existing LHC data in section \ref{sec:pheno}. 
Finally we conclude with a discussion and an outlook.

\section{Model aspects}

We summarize here the main aspects and refer to \cite{Caliri:2024jdk} for further details. The so-called vacuum
misalignment is characterized by an angle $\theta$ which is given by
\begin{align}
f_\pi \sin \theta = v_{\text{SM}} = 246~\text{GeV}\,.
\label{eq:def_fpi}  
\end{align}
with $f_\pi$ being the decay constant of the pNGBs and $v_{\text{SM}}$ is the vacuum expectation value of the
SM Higgs boson. The spin-1 resonances emerge as bound states of the new fermions. Their properties emerge from three types of cosets, $\SU(5)/\SO(5)$, $\SU(4)/\Sp(4)$, and $\SU(4)^2/\SU(4)$.
The resulting spectrum contains a set of vectors $\mathcal V^\mu$ as well as a set of axial-vectors 
$\mathcal A^\mu$, which decay respectively into two or three 
pNGBs\footnote{This is actually an abuse of language. Only in case of the coset SU(3)$\times$SU(3)/SU(3) the names 'vector' and 'axial-vector´ 
coincides with the CP properties of the
spin-1 resonances.}. The latter property originates from 
the symmetric nature of the cosets. We give in table ~\ref{tab:cosets} the resulting pNGBs and spin-1 resonances.
The spin-1 states which do \textit{not} mix with the SM EW vector bosons are indicated by a hat on the corresponding name.
Their masses are given by two independent parameters $M_V$ and $M_A$.

\begin{table}[t]
    \resizebox{\textwidth}{!}{
    \begin{tabular}{|c|ccc|ccc|ccc|}
    \hline
      coset/particles   & \multicolumn{3}{c|}{pNGBs} &  \multicolumn{3}{c|}{$\mathcal A_\mu$} & \multicolumn{3}{c|}{$\mathcal V_\mu$} \\
      &  $\SU(2)^2$
      &  $\SU(2)_D$ & name  &  $\SU(2)^2$
      &  $\SU(2)_D$ & name  &  $\SU(2)^2$
      &  $\SU(2)_D$ & name \\ \hline
    SU(4)/Sp(4)  & (2,2) & 3 & $\varphi$ 
                 & (2,2) & 3 & $a_\mu$ 
                 & (2,2) & 3 & $\hat r_\mu$ \\
                 &  & 1 & $H$
                 &  & 1 & $\hat y_{1\mu}$      
                 &  & 1 & $\hat x_{1\mu}$    \\
    in M8-M9             & (1,1) & 1 & $\eta$ 
                 & (1,1) & 1 & $\hat y_{2\mu}$ 
                 & (3,1)+(1,3) & 3 &  $v_{1\mu}$ \\
                & & & & & & 
          &  & 3 &  $v_{2\mu}$ \\ \hline        
    SU(5)/SO(5)  & (2,2) & 3 & $\varphi$ 
                 & (2,2) & 3 & $a_\mu$ 
                 & (2,2) & 3 & $\hat r_\mu$ \\
                 &  & 1 & $H$
                 &  & 1 & $\hat y_{1\mu}$      
                 &  & 1 & $\hat x_{1\mu}$    \\
    in M1-M7            & (1,1) & 1 & $\eta$ 
                 & (1,1) & 1 & $\hat y_{2\mu}$ 
                & (3,1)+(1,3) & 3 &  $v_{1\mu}$ \\
                 & (3,3) & 5 & $\eta_5 $ 
                & (3,3) & 5 &  $\hat a_{5\mu}$
          &  & 3 &  $v_{2\mu}$ \\       
                & & 3 &  $\eta_3$
                & & 3 &  $\hat a_{3\mu}$
          &  &  &  \\       
                & & 1 &  $\eta_1$
                & & 1 &  $\hat a_{1\mu}$
          &  &  &  \\ \hline        
    SU(4)$^2$/SU(4)  & (2,2) & 3 & $\varphi$ 
                 & (2,2) & 3 & $a_\mu$ 
                 & (2,2) & 3 & $\hat r_\mu$ \\
                 &  & 1 & $H$
                 &  & 1 & $\hat y_{1\mu}$      
                 &  & 1 & $\hat x_{1\mu}$    \\
    in M10-M12              & (2,2) & 3 & $\phi_1$ 
                 & (2,2) & 3 & $\hat a_\mu$ 
                 & (2,2) & 3 & $ r_\mu$ \\
                 &  & 1 & $\phi_2$
                 &  & 1 & $\hat y_{3\mu}$      
                 &  & 1 & $\hat x_{3\mu}$    \\
                 & (1,1) & 1 & $\eta$ 
                 & (1,1) & 1 & $\hat y_{2\mu}$ 
                 & (1,1) & 1 & $\hat x_{2\mu}$ \\
                & (3,1)+(1,3) & 3 &  $\eta_{1}$
                & (3,1)+(1,3) & 3 &  $b_{1\mu}$
          & (3,1)+(1,3) & 3 &  $v_{1\mu}$ \\        
                &  & 3 &  $\eta_{2}$
                &  & 3 &   $b_{2\mu}$ 
          &  & 3 &  $v_{2\mu}$ \\ \hline        
    \end{tabular}
    }
    \caption{List of pNGBs, axial vector and vector states for the three cosets. For each particle
    we give first the $\SU(2)^2 \equiv \SU(2)_L \times \SU(2)_R$ representation, the $\SU(2)_D$ representation and the name used for the latter. 
    Moreover, we list in the first column the models from ref.~\cite{Belyaev:2016ftv} that feature the corresponding coset.}
    \label{tab:cosets}
\end{table}

In the following we focus on  the cosets $\SU(4)/\Sp(4)$ and $\SU(5)/\SO(5)$ as for those we have the 
same mixing patters with the SM vector bosons \cite{Caliri:2024jdk} and refer for the details of the
third coset to \cite{Caliri:2024jdk}. The $W^+$-boson mixes with the states $a^+_\mu$ and a linear
combination of $v_{1\mu}^+$ and $v_{2\mu}^+$. We denote the corresponding mass eigenstate by $V^+_{1\mu}$ in the following. In the neutral sector the EW vector bosons mix with $a^0_\mu$, $v_{1\mu}^0$ and $v_{2\mu}^0)$.
We denote the mass eigenstates which are dominantly a linear combination of $v_{1\mu}^0$ and $v_{2\mu}^0$ by
$V^0_{1\mu}$ and $V^0_{2\mu}$. We note, that the mixing with the axial states $a^+_\mu$ and $a_{\mu}^0$
vanishes in the limit $\sin\theta\to 0$.

The corresponding mass matrices are diagonalized by orthogonal
rotation matrices which we denote by $\mathcal{C}$ and $\mathcal{N}$ 
for the charged and neutral sectors, respectively: 
\begin{align}
   \left( \tilde W^+_\mu , a^+_\mu , v_{1\mu}^+ , v_{2\mu}^+ \right)^T &= \mathcal{C} 
   \left( W^+_\mu , A^+_\mu , V_{1\mu}^+ , V_{2\mu}^+  \right)^T = \mathcal{C} R_\mu^+ \,,  \\
   \left( B_\mu , \tilde W^3_\mu, a^0_\mu, v_{1\mu}^0, v_{2\mu}^0 \right)^T 
   &= \mathcal{N} \left( A_\mu, Z_\mu , A^0_\mu , V_{1\mu}^0 , V_{2\mu}^0 \right)^T = 
   \mathcal{N} R_\mu^0 \,, \label{eq:mass-eigenstates-basis}
\end{align}
denoting the mass eigenstate vectors by $R_\mu^+$
and $R_\mu^0$.

We now collect the interactions that facilitate either the production or the decay of the heavy spin-1 resonances. We focus on the states $V_1^0, V_2^0, V_1^+$ that still mix with SM vector bosons 
even if $\sin\theta \to 0$. This mixing generates couplings between these spin-1 resonances and SM fermions: 
\begin{align}
    \mathcal{L}_\mathrm{CC} &= \frac{\hat g}{\sqrt{2}} \sum_{i,f,f'} \mathcal{C}_{1i} \Bar{\psi}_f \slashed{R}^+_i P_\mathrm{L} (V_{\tiny \mathrm{CKM}})_{ff'}\psi_{f'} + \hc \,, \label{eq:LagCC}\\
    \mathcal{L}_\mathrm{NC} &=  \sum_{i,f} \Bar{\psi}_f \slashed{R}^0_i \left( g_{\mathrm{L}i}^f P_\mathrm{L} + g_{\mathrm{R}i}^f P_\mathrm{R} \right) \psi_f  \,,\label{eq:LagNC} \\
\text{ with } \quad &
g_{\mathrm{L}i}^f = \hat g T^3_f \mathcal{N}_{2i} + \hat g' Y_{fL} \mathcal{N}_{1i}  \quad \text{ and } \quad
g_{\mathrm{R}i}^f = \hat g' Y_{fR} \mathcal{N}_{1i}.
\end{align}
Here $T^3_f$ is the weak isospin of the fermion $f$ and $Y_{fL,fR}$ are the corresponding hypercharges. 
These couplings give rise to single production as discussed below.

The third generation quarks get an additional contribution from the mixing between the elementary fields and the top partners in models with partial compositeness (PC), which we parametrize as
\begin{align}\label{eq:simplifiedmodelVtt}
    \mathcal{L}_\mathrm{PC} &= \Bar{t} \left(\slashed{V}^0_{1}+\slashed{V}^0_{2} \right)  \left( g_{t,L} P_L + g_{t,R} P_R \right) t + \Bar{b} \left(\slashed{V}^0_{1}+\slashed{V}^0_{2} \right)  \left( g_{b,L} P_L  \right) b +  g_{tb,L}
    \Bar{t} \slashed{V}^+_{1}  P_L b \,.
\end{align}
Due to the small mixing of the bottom quark with its partner the $g_{b,L}$ will be small.
Moreover,  we assume here for simplicity that the couplings of $V^0_1$ and $V^0_2$ are the
same. In practice they differ slightly due to the difference in the corresponding entries
of $\mathcal{N}$, which are however small at the numerical level.

The mixing between the spin-1 resonances and the SM vector boson induce 
couplings of one spin-1 resonance to two electroweak vector bosons which originate 
from the terms
\begin{align}
    \mathcal{L} \supset -\mathrm{i} &\Big( \hat g \tilde W^{+\nu} \tilde W_\mu^- \partial^\mu \tilde W_\nu^3 + \frac{\gt}{\sqrt{2}} \left( (a^{+\nu} v^-_{1\mu} + v^{+\nu}_1 a^-_\mu) \partial^\mu a_\nu^0 + (v^{+\nu}_1 v^-_{2\mu} + v^{+\nu}_2 v^-_{1\mu}) \partial^\mu v_{2\nu}^0 \right) \nonumber \\
    &+ \frac{\gt}{\sqrt{2}} \left(a^{+\nu} a^-_\mu + v^{+\nu}_1 v^-_{1\mu} + v_2^{+\nu} v^-_{2\mu} \right) \partial^\mu v_{1\nu}^0  \Big) + \text{permutations } \,.
    \label{eq:int_3v}
\end{align}
Moreover, there are couplings of a spin-1 resonance to two pNGBs which is characterized by the
vector-pNGB-pNGB coupling constant $g_{V \pi \pi}$ which is the analog of $g_{\rho\pi\pi}$ coupling in QCD. There is a second contribution to this coupling originating from mixing of the spin-1 resonances
with the SM vector bosons \cite{Caliri:2024jdk}. Moreover, the spin-1 resonances couple to the 
Higgs boson and one SM vector boson.

The couplings and mixing matrices are characterized by four parameters: the vector mass parameter 
$M_V$, the ratio of axial to vector mass $\xi=M_A/M_V$,  the coupling scale of vectors to two
pNGBs $g_{V\pi\pi}$ and the decay constant $f_\pi$. In addition one has the strong coupling 
$\gt$ of the new sector as a free parameter.
In the following we will fix the pion decay constant $f_\pi$ to $1\,\mathrm{TeV}$ as its
variation only mildly affects the decay channels of interest.
Both, lattice studies \cite{Ayyar:2018zuk,Bennett:2019jzz,Bennett:2019cxd,Bennett:2023qwx} 
and holographic models using gauge/gravity duality \cite{Erdmenger:2020lvq,Erdmenger:2020flu,Elander:2020nyd,Elander:2021bmt,Erdmenger:2023hkl,Erdmenger:2024dxf} yield $\xi > 1$ and, thus, 
we set $\xi=1.4$ in the following.
Additionally, we use the SM values of the electric charge $e$ and mass of the $Z$ boson $M_Z$ as input parameters.

\section{Relevant phenomenological aspects}
%\label{sec:pheno}

\begin{figure}[t]
    \centering
    \includegraphics[width=0.5\linewidth]{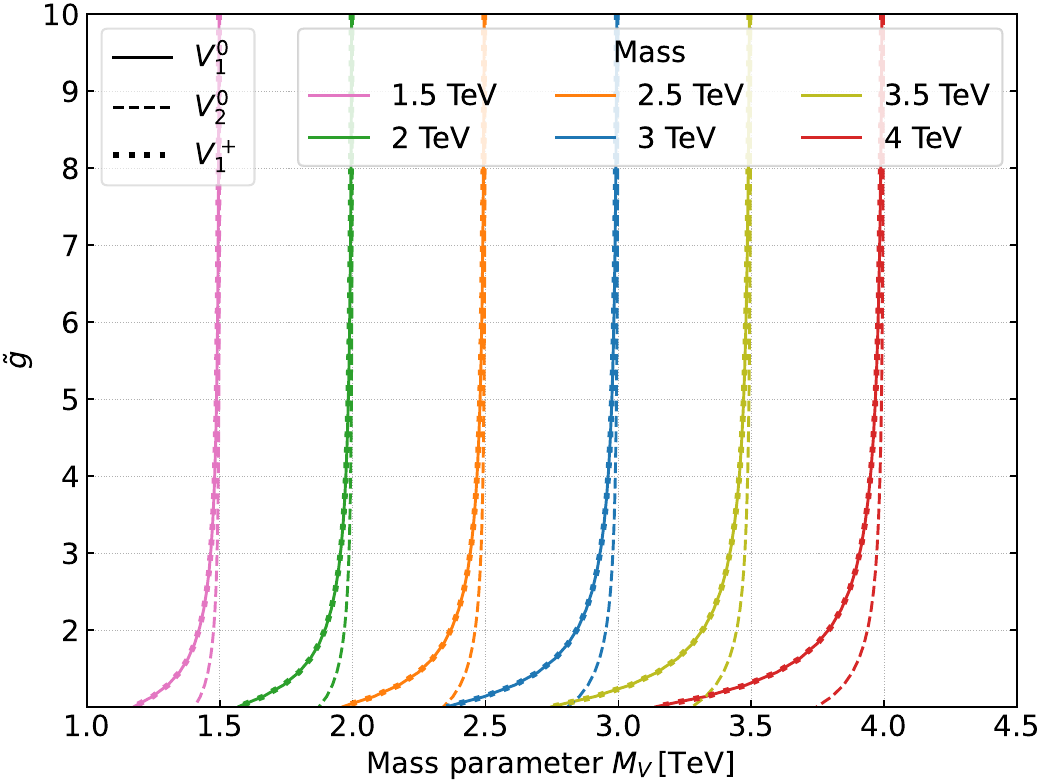}
    \caption{Contour lines for the masses of $V^+_{1\mu}$, $V^0_{1\mu}$ and $V^0_{2\mu}$ in the $M_V$-$\gt$ plane. The results look nearly identical for each coset $\SU(4)/\Sp(4)$, $\SU(5)/\SO(5)$ and $\SU(4)\times \SU(4)/\SU(4)$.}
    \label{fig:mass-contour-plot}
\end{figure}

The states, which mix with
the SM electroweak bosons even in the limit
$\sin\theta \to 0$,  can be singly produced with a sizeable cross section at the LHC as we will see below. They are denoted as $V^+_1$, $V^0_1$ and $V^0_2$.
The first two states stem essentially from $(3,1)$ of $\SU(2)_L\times \SU(2)_R$ whereas $V^0_2$ is mainly the neutral state of $(1,3)$ 
mixing primarily with the hypercharge boson. This is also visible in fig.~\ref{fig:mass-contour-plot} 
where we show corresponding contour lines for the  masses of these states in the $M_V$-$\gt$ plane.
All states are nearly mass degenerate for $\gt \gsim 4$.

We are interested in the bounds on these states from existing LHC. For these considerations, it is 
useful to group the various decay channels as follows
\begin{alignat}{3}
    &\mathcal V^0 \to q\bar q, \, l^+ l^-, \,\nu\bar \nu, \qquad\qquad &&\mathcal V^0 \to t\bar t, \qquad\qquad &&\mathcal V^0 \to \pi\pi,\, HZ,\, W^+ W^-, \\
    &\mathcal V^+ \to q\bar q', \, l^+ \nu, \qquad &&\mathcal V^+ \to t\bar b, \qquad &&\mathcal V^+\to \pi\pi , \, W^+ Z,\, W^+ H.
\end{alignat}
The phenomenology of these states depends on various unknown parameters. Thus, we consider four exemplary scenarios which are characterized by combinations of couplings to pNGBs and the top quark:
\begin{alignat}{4}
    &\mathbf{SM}\, \boldsymbol t: \qquad &&g_{t,L/R} = g_{Ztt,L/R}^\mathrm{SM}, \qquad \qquad  \qquad &&g_{b,L} = g_{Zbb,L}^\mathrm{SM}, \qquad \qquad  &&g_{tb,L} = g_{Wtb}^\mathrm{SM}\,, \label{eq:SMtt} \\
    &\textbf{PC}\, \boldsymbol t: &&g_{t,L}=\frac{1}{\sqrt{10}}\,,\, g_{t,R}=\frac{3}{\sqrt{10}}\,,&&g_{b,L}=\frac{1}{\sqrt{10}}\,,  && g_{tb,L}=\frac{1}{\sqrt{5}}\,. \label{eq:PCtt} 
\end{alignat}
For the pNGB couplings we consider
\begin{alignat}{2}
    &\mathbf {weak}\, \boldsymbol \pi: \qquad &&g_{V\pi\pi} = 0\,, \\
    &\mathbf {strong}\, \boldsymbol \pi: \qquad &&g_{V\pi\pi} = 4 \,.
\end{alignat}

\begin{figure}[t]
    \centering
    \begin{subfigure}{.32\textwidth}
        \includegraphics[width=\linewidth]{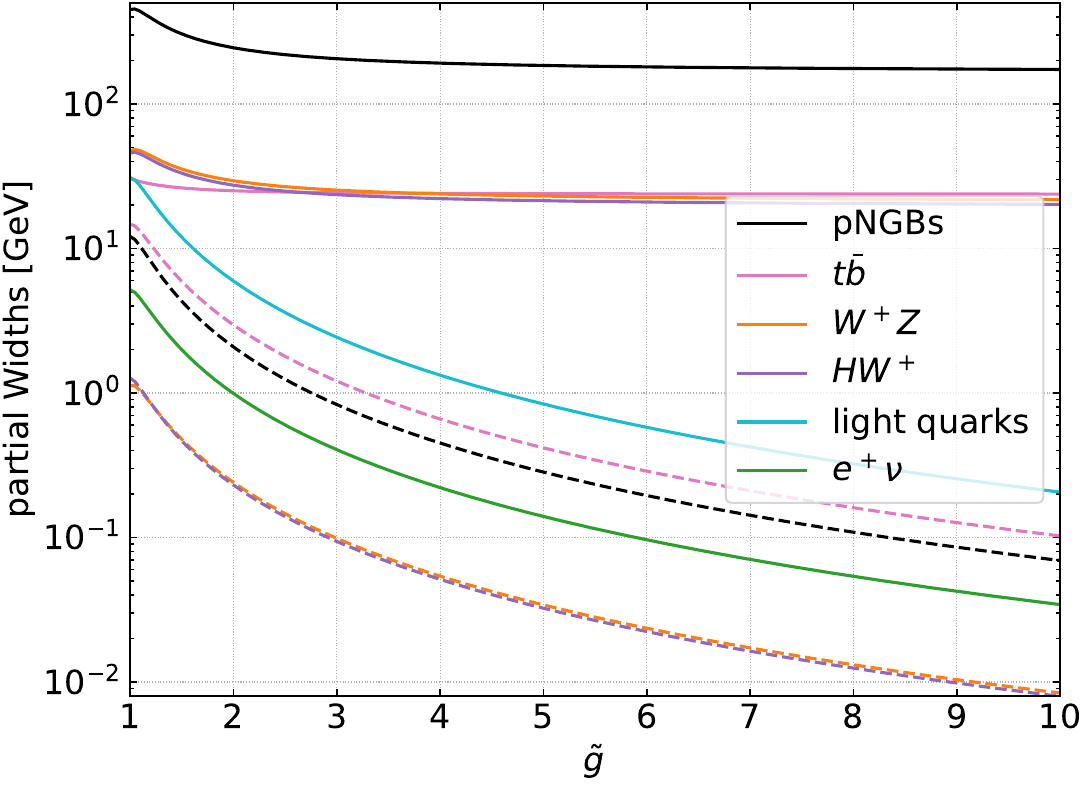}
        \caption{partial widths of $V_{1\mu}^+$}
    \end{subfigure}
    \begin{subfigure}{.32\textwidth}
        \includegraphics[width=\linewidth]{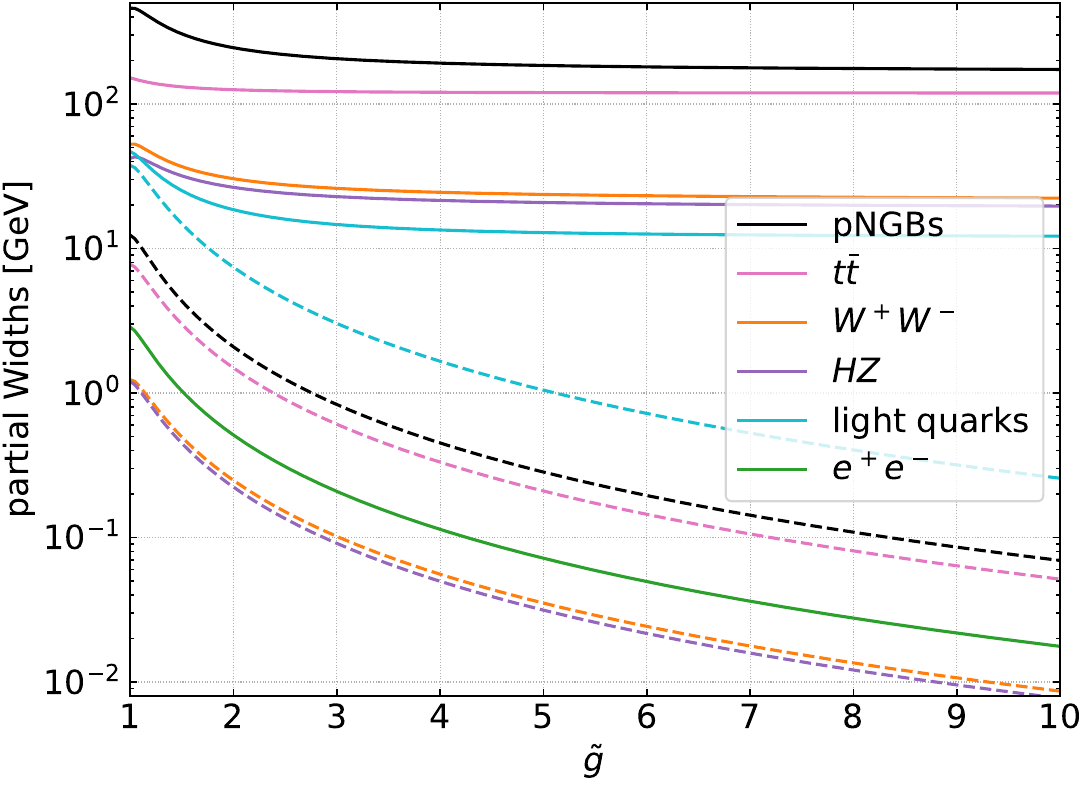}
        \caption{partial widths of $V_{1\mu}^0$}
    \end{subfigure}
    \begin{subfigure}{.32\textwidth}
        \includegraphics[width=\linewidth]{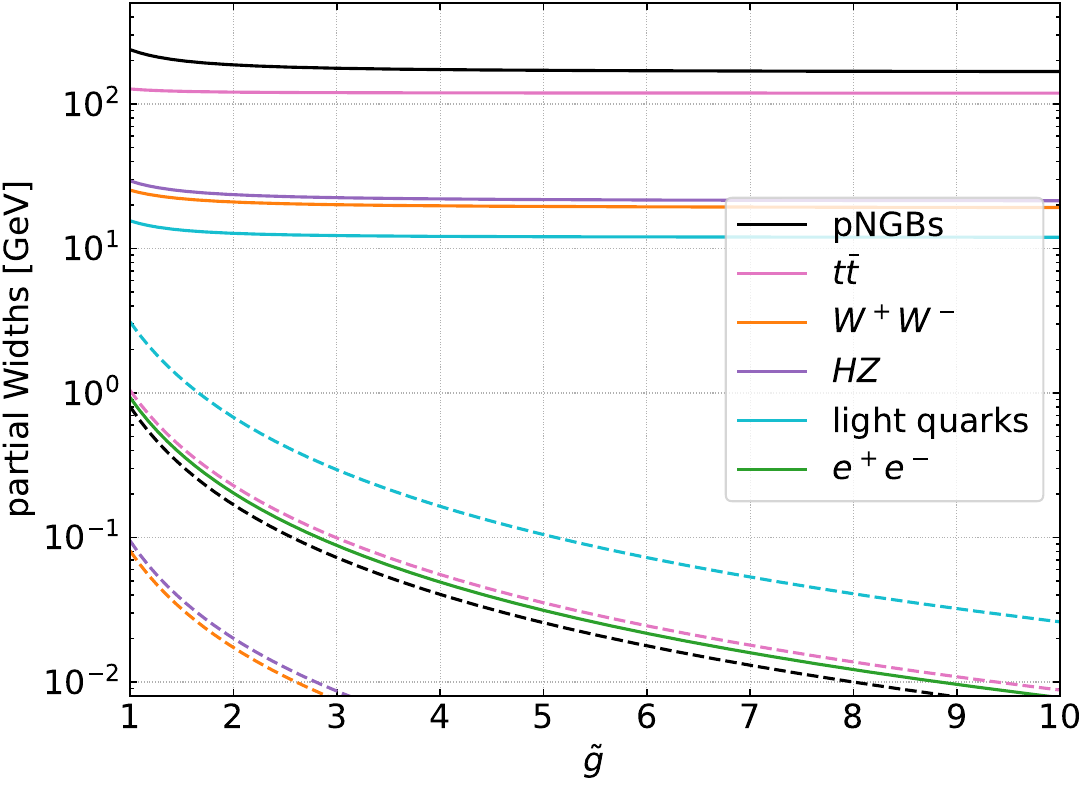}
        \caption{partial widths of $V_{2\mu}^0$}
    \end{subfigure}
    \caption{Partial decay widths of selected spin-1 resonances for the $\SU(5) / \SO(5)$ coset. 
    We have set   $M_V=3000$~GeV and $M_\pi=700$~GeV. 
  pNGB, $W^+ W^-$, $HZ$, $W^+Z$ and $H W^+$  channels: the solid lines of the correspond to $\gV=4$
   and the dashed lines correspond to $\gV=0$. Top quark channels: the solid lines correspond to 
    $g_t = 1$ and the dashed lines to SM-like couplings. 
  These also represent the partial widths for the $\SU(4)/\Sp(4)$ coset for which, however, the black lines (additional pNGB channels) are absent.
}
  \label{fig:partial-widths}
\end{figure}

We show in fig.~\ref{fig:partial-widths}  the partial decay widths for the various scenarios. for the 
For the $\SU(4)/\Sp(4)$ coset, the black lines representing the decays into the additional pNGBs are
absent as there is no coupling of the gauge singlet $\eta$ to any combination of the electroweak
vector bosons and any of the considered spin-1 resonances. We have set the pNGB mass to 700~GeV such 
that existing LHC bounds are satisfied \cite{Cacciapaglia:2022bax} and $M_V = 3$~TeV.
We show the widths for the decays into two bosons for the cases $g_{V\pi\pi} = 4$ and 0 as solid 
and dashed lines, respectively. Similarely, for the top quark channel we distinguish the PC $t$ and 
the SM $t$ scenarios by solid and dashed lines.
Important features to be observed:
\begin{itemize}
 \item $\textbf{PC}\, \boldsymbol t,\, \mathbf{strong}\, \boldsymbol \pi$: In scenarios with large
   $g_{V\pi\pi}$ and large couplings to top-quarks the \mbox{spin-1} resonances decay dominantly
   into the additional pNGBs and $t\bar{t}$ followed by decays into $HV$ and $WV$ ($V=Z,W$) 
   in case the $\SU(5)/\SO(5)$ coset is realized. In case of the $\SU(4)/\Sp(4)$ coset, the dominant 
   channel will be $t\bar{t}$ followed by $HV$ and $WV$. The enhancement of the
    $HV$ channel is due to the longitudinal components of the vector bosons.   
 \item $\textbf{PC}\, \boldsymbol t,\, \mathbf{weak}\, \boldsymbol \pi$: In scenarios with large
   additional top Yukawa couplings and $g_{V\pi\pi} \lsim \mathcal O(0.1)$, the $t\bar{t}$ channel
    dominate independent of the coset.     
    \item $\textbf{SM}\, \boldsymbol t,\, \mathbf{strong}\, \boldsymbol \pi$: for large  
    $g_{V\pi\pi}$ couplings and SM-like couplings to top quarks, the decays into the additional pNGBs
    dominate followed by the $HV$ and $WV$ channels in case of the $\SU(5)/\SO(5)$ coset whereas 
    the latter tow channel dominate in case of $\SU(4)/\Sp(4)$. 
    \item $\textbf{SM}\, \boldsymbol t,\, \mathbf{weak}\, \boldsymbol \pi$: In case that all
     additional couplings are small, the decay patterns are similar to those of $W$ and $Z$ bosons 
     but for the additional decays into top quarks. In case of $\SU(5)/\SO(5)$, the decays into 
     the additional pNGBs are rather important.
\end{itemize}
These observations imply that we have to take into account the cascade decays via intermediate
pNGBs. Here we summarize the various possibilities and refer to the literature for further
details \cite{Agugliaro:2018vsu,Cacciapaglia:2022bax,Ferretti:2016upr}. We assume in the following
that none of the additional states mixes with the Higgs boson. We denote here the additional states
generically as $S^0$, $S^+$ and $S^{++}$ depending on their charge. 

As one limiting case we consider the case that the pNGBs decay dominantly into third generation 
quarks in all cosets and dub this the fermiophilic scenario below. These decays are mainly induced
by the mixing of the top and bottom quarks with the top-partners. 
The neutral states $S^0$ decay as
\begin{align}
S^0_i \to t \bar{t}\,,\quad b \bar{b}     
\end{align}
with the $b \bar{b}$ channel being suppressed by the ratio $(m_b/m_t)^2$. 
 Similarly, $S^+$  decays as
\begin{align}
    S^+ \to t \bar{b}.
\end{align}
The coset SU(5)/SO(5) features a doubly charged scalar which decays as
\begin{align}
    S^{++} \to W^+ t \bar{b}
\end{align}
via an intermediate $S^+$ \cite{Cacciapaglia:2022bax}.

In case that these pNGB couplings to quarks are absent -- dubbed fermiophobic scenario --,
decays into two SM vector bosons induced by the anomalous WZW terms are relevant. 
Cascade decays into a vector boson and another pNGB are also important in scenarios with a mass
splitting between the pNGBs  \cite{Cacciapaglia:2022bax}. 
We take here the SU(5)/SO(5) coset as a prototypical example where all pNGBs but $\eta_3^0$ 
have anomaly induced couplings. In scenarios in which
the triplet is the lightest state of in scenarios with mass degenerate additional pNGBs, which we 
assume in the following, the CP-even $\eta_3^0$ undergoes only three-body decays via an 
off-shell pNGB:
\begin{align}
    \eta_3^0 &\to W^\mp {\eta^\pm_{3,5}}^* \to W^+ W^- \gamma,\,W^+ W^-Z \\
    \eta_3^0 &\to Z {\eta^0_{1,5}}^* \to Z Z Z ,\, Z Z \gamma ,\, Z \gamma \gamma
    \,.
\end{align}
Their analytic expressions of the corresponding partial widths are given in \cite{Caliri:2024jdk}.

\section{Constraints due to existing LHC data}
 \label{sec:pheno}

\begin{figure}[t]
    \centering
    \begin{subfigure}{0.33\linewidth}
        \includegraphics[width=\linewidth]{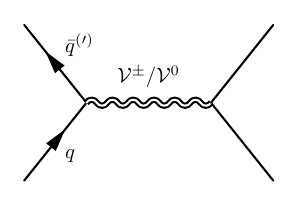}
        \vspace{5ex}
    \end{subfigure} \quad
    \begin{subfigure}{0.50\linewidth}
        \includegraphics[width=\linewidth]{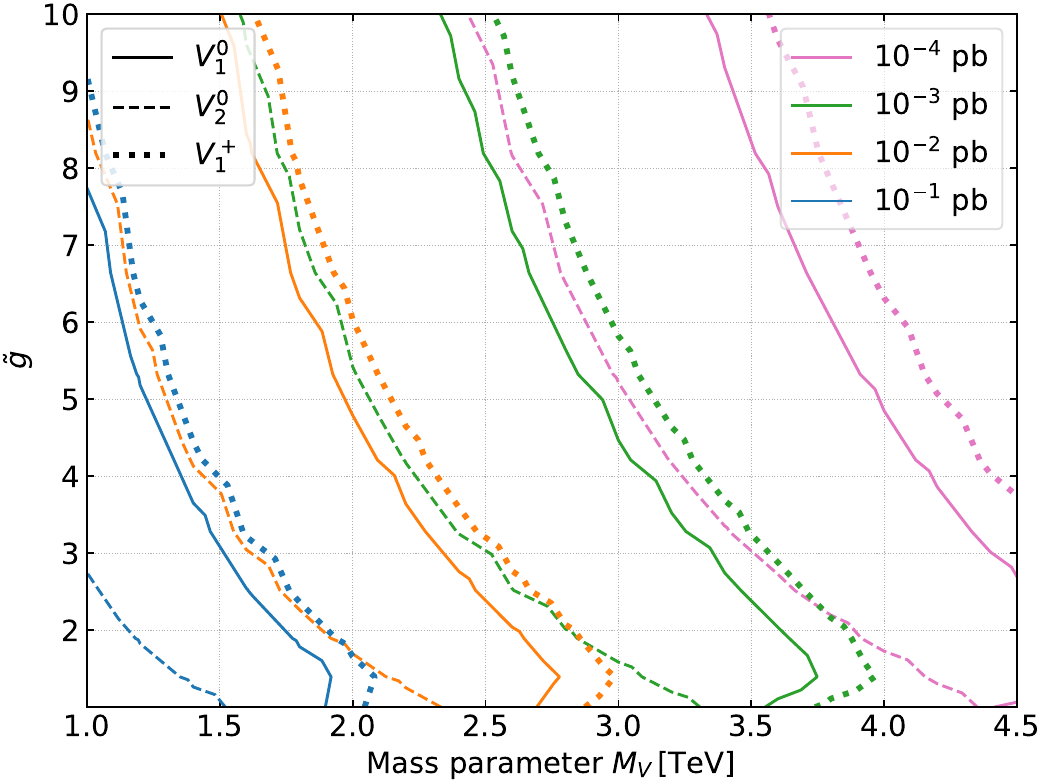}
    \end{subfigure}
    \caption{Drell-Yan production of heavy vectors. To the left: typical Feynman diagrams.
    To the right: production cross sections at $\sqrt s = 13$~TeV of the heavy vector states in the
     $M_V$-$\tilde g$-plane for a small $g_{V\pi\pi}$ coupling and nearly SM-like couplings to 
     top-quarks. The cross sections are the same for both cosets, $\SU(5)/\SO(5)$ and $\SU(4)/\Sp(4)$.}
    \label{fig:feynman-dy}
\end{figure}

The states $V^+_1$  and $V^0_{1,2}$ have sizeable couplings to quarks of the first two generations
due to the mixing of these states with the SM vector bosons.
Thus, they can be singly produced at the LHC as exemplary shown in fig.~\ref{fig:feynman-dy} for 
the tow cosets considered. The various decay modes of the spin-1 resonances discussed above lead to
to multiple signatures that have been searched for at the LHC. In particular the following searches
are relevant  to constrain the parameter space of our models considered:
\begin{itemize}
    \item an ATLAS search for $Z' \to \ell^+ \ell^-$ using 139~fb$^{-1}$ \cite{ATLAS:2019erb},
    \item an ATLAS search for $Z' \to t\bar t$ using 139~fb$^{-1}$ \cite{ATLAS:2020lks},
    \item an ATLAS search for $W' \to \ell^+ \nu$ using 139~fb$^{-1}$ \cite{ATLAS:2019lsy},
    \item an ATLAS search for $W' \to t\bar b$ using 139~fb$^{-1}$ \cite{ATLAS:2023ibb}.
\end{itemize}
To this end, all relevant vertices have been implemented in the \texttt{FeynRules} \cite{Alloul:2013bka} format to obtain an UFO library \cite{Degrande:2011ua}. This UFO has been used to generate
events of the respective process at $\sqrt s = 13$~TeV with \texttt{MadGraph5\_aMC@NLO} \cite{Alwall:2014hca} v3.5.3. Dynamical renormalization and factorization scales have been used together with the \texttt{NNPDF~2.3} set of parton distribution functions \cite{Ball:2012cx} implemented in \texttt{LHAPDF} \cite{Buckley:2014ana}. This has been used to calculate the cross sections of a given process for a grid of parameter points, which are compared  to the upper limits obtained 
from the above listed searches to derive exclusion limits in the $M_V$-$\tilde g$-plane. 

We have recast searches for final states with two bosons as they are
not covered by any experimental search. To this end, we have first showered and hadronized the events with \texttt{Pythia8} \cite{Sjostrand:2014zea} to produce a \texttt{HepMC} file \cite{Dobbs:2001ck}.
The hadronized events have passed to \texttt{MadAnalysis5} \cite{Conte:2012fm,Conte:2014zja,Dumont:2014tja,Conte:2018vmg} 
v1.10.9beta and \texttt{CheckMATE} \cite{Drees:2013wra,Dercks:2016npn} commit number \texttt{1cb3f7}. 
Both tools cluster the jets with the anti-$k_T$ algorithm \cite{Cacciari:2008gp} implemented in the 
\texttt{FastJet} library \cite{Cacciari:2011ma} and simulate the detector response with 
\texttt{Delphes 3} \cite{deFavereau:2013fsa}.
These events have then been run through the kinematic cuts of the recast searches, and from the 
number of remaining events an exclusion value has been calculated with the CL$_s$ method 
\cite{Read:2002hq} for every signal region.
We have collected the observed exclusion for the signal region for every search that had the strongest 
expected bound. In addition, we have run the events against the SM measurements implemented in 
\texttt{Rivet} \cite{Bierlich:2019rhm} v3.1.8 and we have evaluated the results with \texttt{Contur}
\cite{CONTUR:2021qmv} v2.4.4. We report the strongest exclusion from any individual search as 
final result. Note, that we have not performed any statistical combination beyond what is 
implemented in these tools.
From this we have obtained contours of the exclusion at 95\% CL as bound in the 
$M_V$-$\tilde g$-plane. We note for completeness, that the regions with small 
$\tilde g \lesssim 2$ are not entirely reliable for the scenarios with strong pNGB coupling as 
the widths of the vector resonances $V_{1\mu}^{0,\pm}$ exceed $10\%$ of their respective mass.

\begin{figure}[p]
    \centering
    \begin{subfigure}{0.47\linewidth}
        \includegraphics[width=\linewidth]{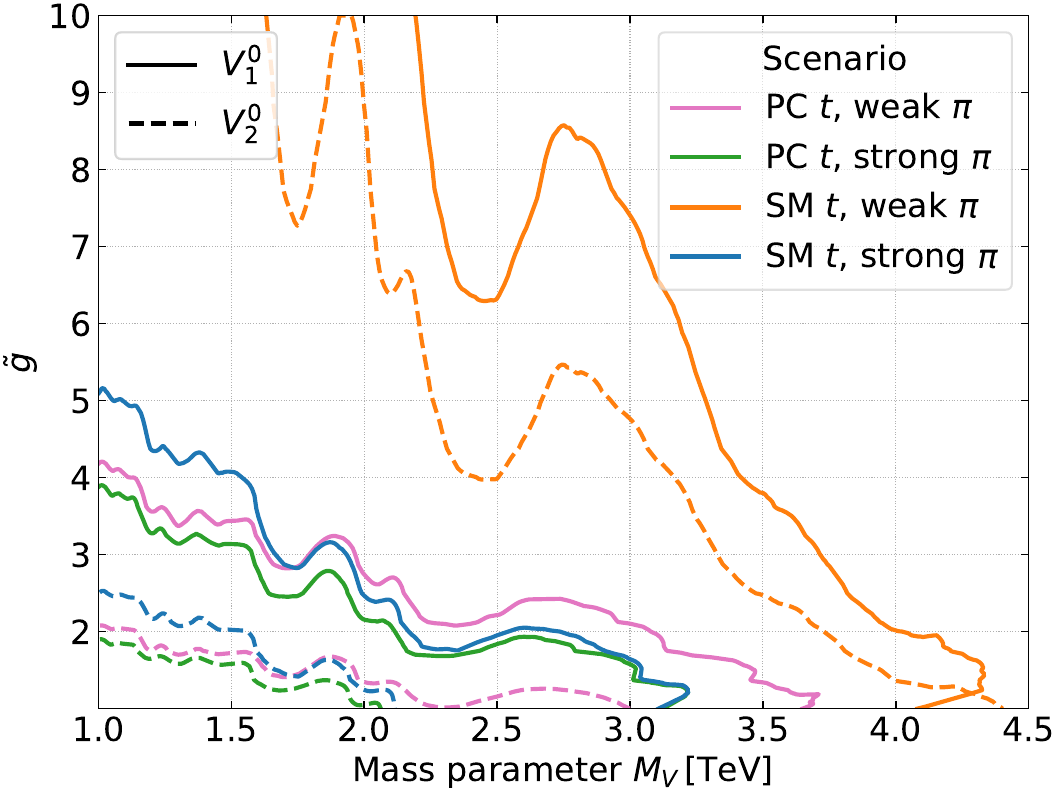}
        \caption{Bounds from $\mathcal V^0\to \ell^+ \ell^-$}
        \label{fig:bounds_su5_ll}
    \end{subfigure}\quad
    \begin{subfigure}{0.47\linewidth}
        \includegraphics[width=\linewidth]{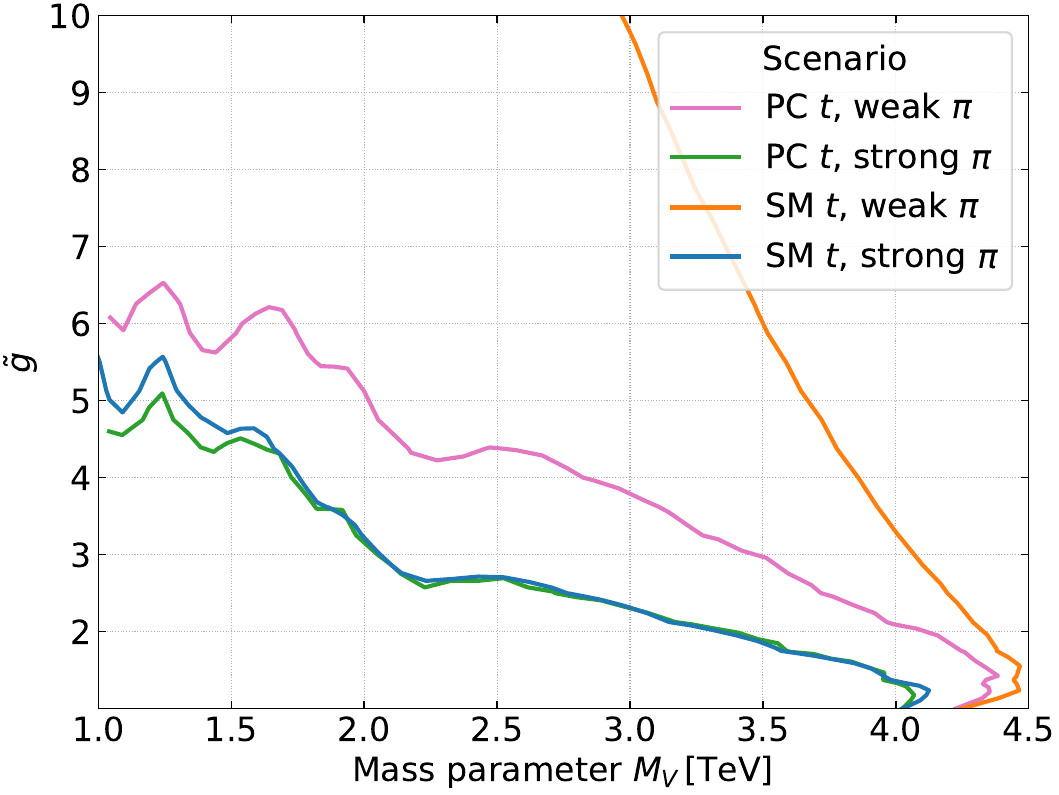}
        \caption{Bounds from $\mathcal V^\pm \to \ell^\pm \nu$}
        \label{fig:bounds_su5_lv}
    \end{subfigure}\vspace{1ex}

    \begin{subfigure}{0.47\linewidth}
        \includegraphics[width=\linewidth]{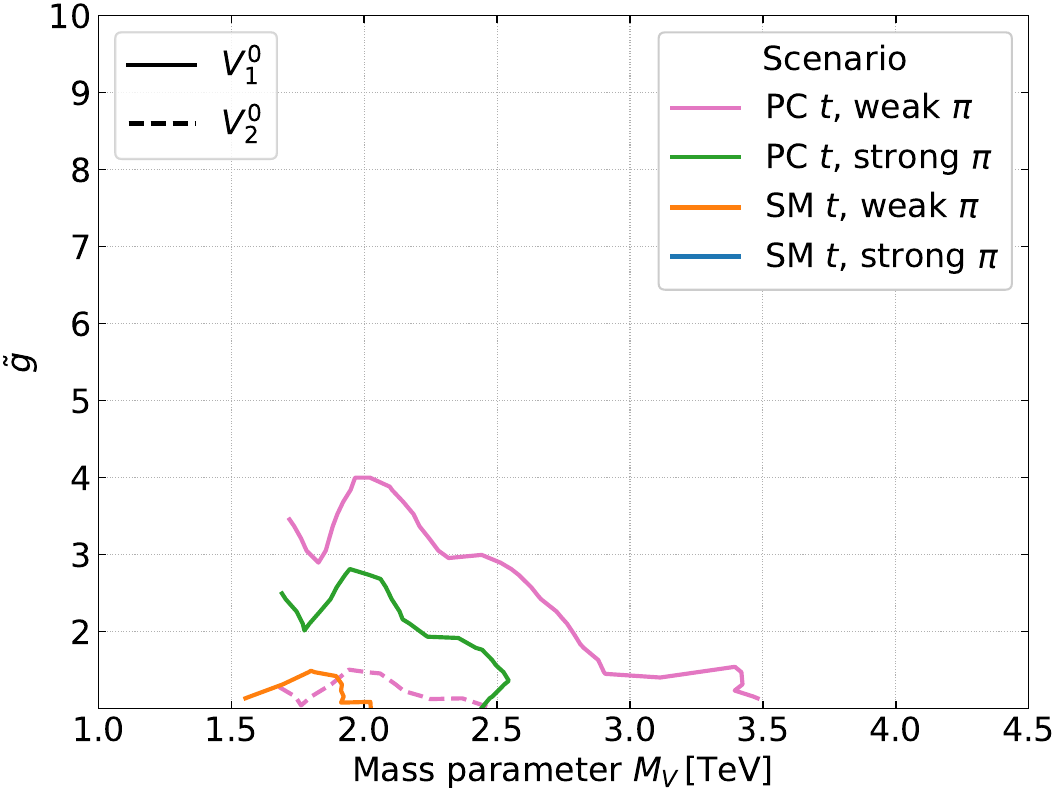}
        \caption{Bounds from $\mathcal V^0 \to t\bar t$}
        \label{fig:bounds_su5_tt}
    \end{subfigure}\quad
    \begin{subfigure}{0.47\linewidth}
        \includegraphics[width=\linewidth]{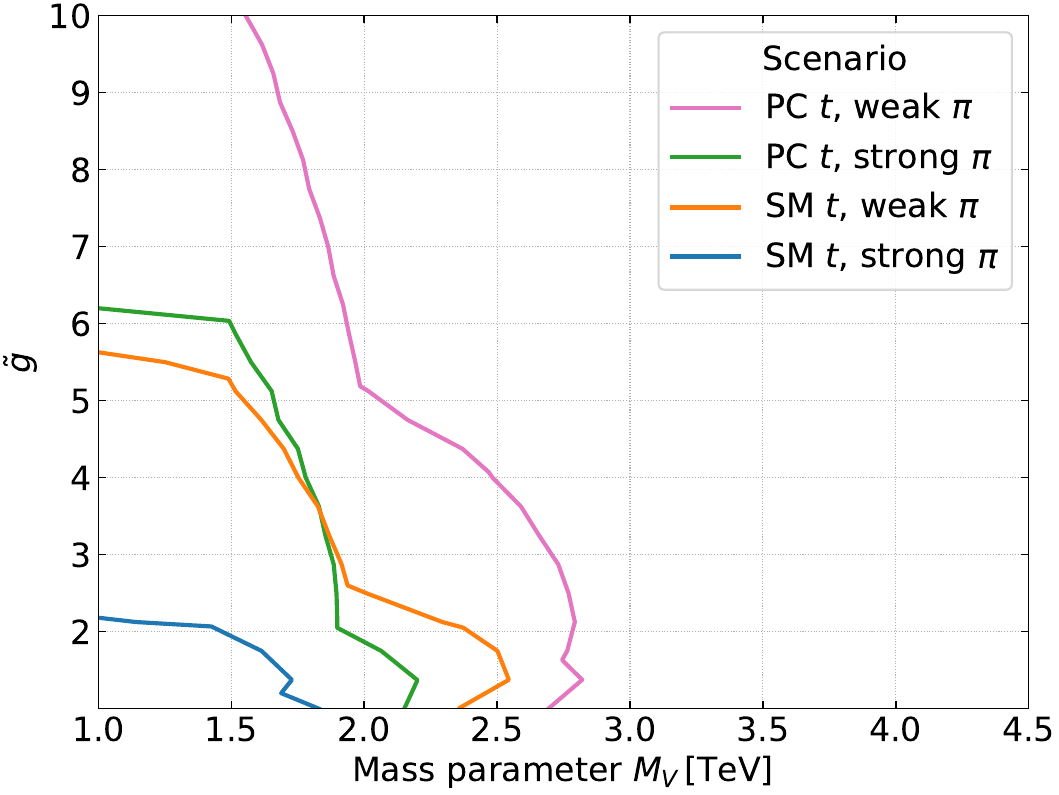}
        \caption{Bounds from $\mathcal V^\pm \to tb$}
        \label{fig:bounds_su5_tb}
    \end{subfigure}

    \begin{subfigure}{0.47\linewidth}
        \includegraphics[width=\linewidth]{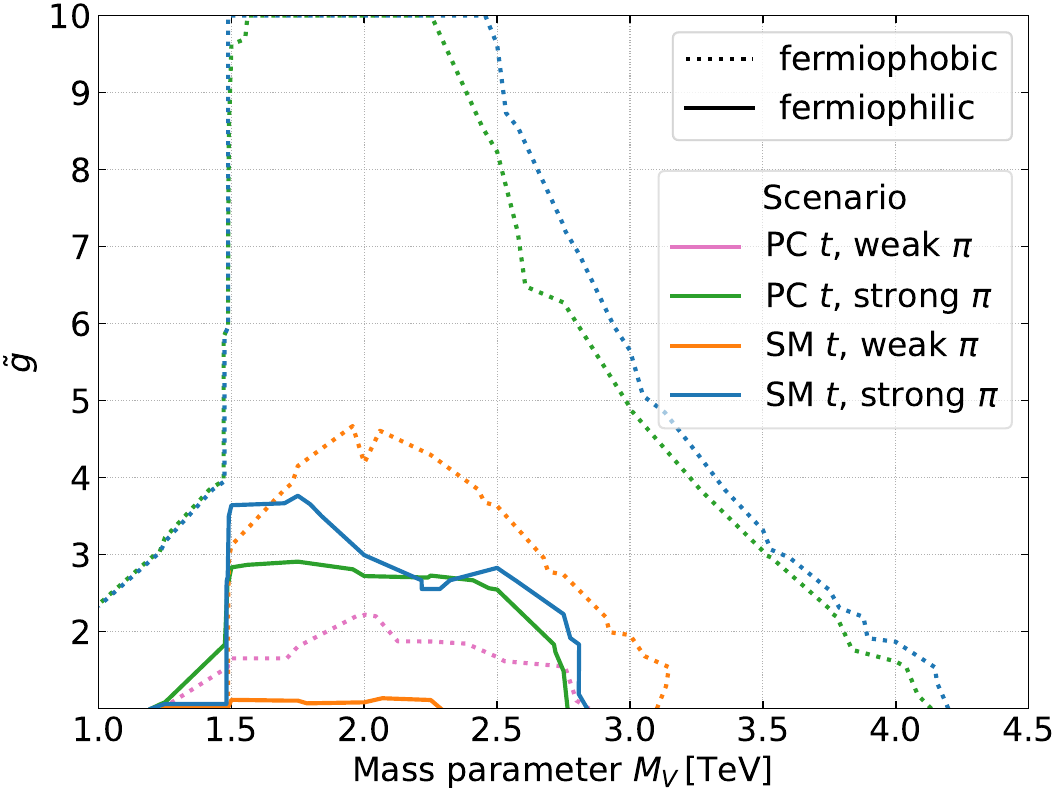}
        \caption{Bounds from $\mathcal V\to \pi\pi$}
        \label{fig:bounds_su5_pipi}
    \end{subfigure}\quad
    \begin{subfigure}{0.47\linewidth}
        \includegraphics[width=\linewidth]{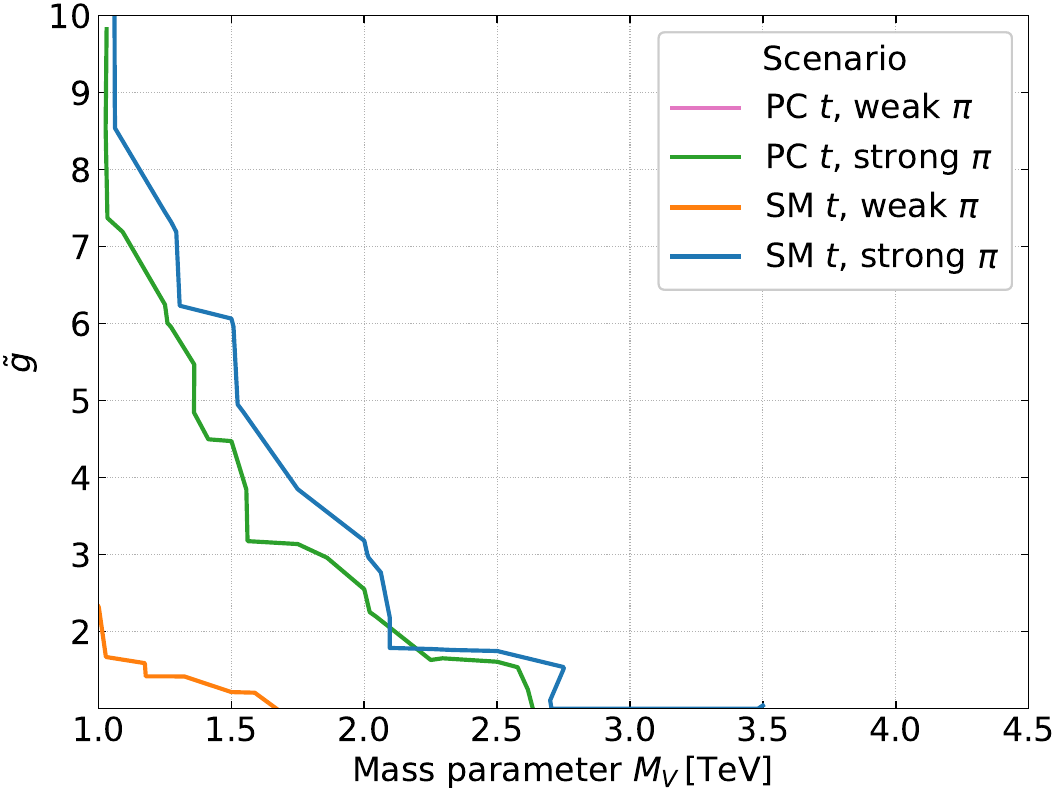}
        \caption{Bounds from $\mathcal V \to H Z, H W^\pm, W^+ W^-, W^\pm Z$}
        \label{fig:bounds_su5_higgs}
    \end{subfigure}
    \caption{Bounds on the single production of heavy spin-1 resonances in the $\SU(5)/\SO(5)$ coset
     for a pNGB mass of 700~GeV. 
    In the scenarios, ``SM $t$'' means the couplings of the $\mathcal V^0/ \mathcal V^\pm$ to $tt/tb$ 
    are given as in eq.~\ref{eq:SMtt} whereas for ``PC $t$'' they are set to the values in
     eq.~\ref{eq:PCtt}. In case of pNGBs, ``weak'' and ``strong $\pi$'' refers to $g_{V\pi\pi}=0$ 
     and $g_{V\pi\pi}=4$, respectively. In plots (a)-(d) the upper limits on the cross sections 
     are obtained from direct searches \cite{ATLAS:2019erb,ATLAS:2020lks,ATLAS:2019lsy,ATLAS:2023ibb}.
    In plot (e) we distinguish between fermiophobic and fermiophilic scenarios of the pNGBs as 
    discussed in the text. The bounds are derived from recasts of \cite{ATLAS:2018nud} and \cite{ATLAS:2021twp,ATLAS:2021fbt,CMS:2019zmd,CMS:2017abv}, respectively.
    The bounds in plot (f) are derived from recasts of \cite{CMS:2017moi,CMS:2019xjf,Mrowietz:2020ztq,CMS:2019xud,Conte:2021xtt,CMS:2019zmn,ATLAS:2020wzf,ATLAS:2022nrp,ATLAS:2021jgw,CMS:2022ubq,ATLAS:2019zci,ATLAS:2019rob}.
    The regions with small $\tilde g \lesssim 2$ have to be taken with a grain of salt for 
    scenarios with strong $\pi$ since the resonances are no longer narrow.}
    \label{fig:allbounds_su5so5}
\end{figure}
We show in fig.~\ref{fig:allbounds_su5so5} the resulting bounds are shown in  for the coset
$\SU(5)/\SO(5)$ as an example and refer to \cite{Caliri:2024jdk} for the other two. Note, that
the x-axis is not the physical mass of the vectors but the mass parameter $M_V$.
We see in figs.~\ref{fig:bounds_su5_ll} and \ref{fig:bounds_su5_lv}, that strong bounds are obtained if
both the couplings to top quarks and pNGBs are small, leaving a sizable branching ratio for decays 
into leptons.
In the other three scenarios the bounds are similar to each other.
The bounds from $\mathcal V^\pm \to t b$ are significantly stronger
compared to the bounds from the decays of the neutral resonances into $t\bar{t}$, see 
figs.~\ref{fig:bounds_su5_tt} and \ref{fig:bounds_su5_tb}, due to the increased cross section of 
the charged channel. 

We show the bounds from decays into pNGBs in the fermiophilic (solid lines) and fermiophobic 
(dotted lines) scenarios in fig.~\ref{fig:bounds_su5_pipi}. The latter are strongly constrained by the 
recast of ref.~\cite{ATLAS:2018nud}, a search for photonic signatures of supersymmetry implemented 
in \texttt{CheckMATE}. The bounds are derived for a common pNGB mass $M_\pi = 700$~GeV to evade constraints from Drell-Yan production of pNGBs~\cite{Cacciapaglia:2022bax}. The sudden drop of the
exclusion lines at $M_V \approx 2 M_\pi$ is due to kinematic suppression of the pNGB channels.
The bounds on pNGB decays into quarks are considerably weaker.
In figure~\ref{fig:bounds_su5_higgs} we show the bounds derived from decays into two gauge bosons or
one gauge and one Higgs boson. For small masses, these channels dominate in the 
strong\,$\boldsymbol{\pi}$ scenarios, but get strongly suppressed above the threshold 
for pNGB pair production.

\begin{figure}[t]
    \centering
    \begin{subfigure}{0.47\linewidth}
     \includegraphics[width=\linewidth]{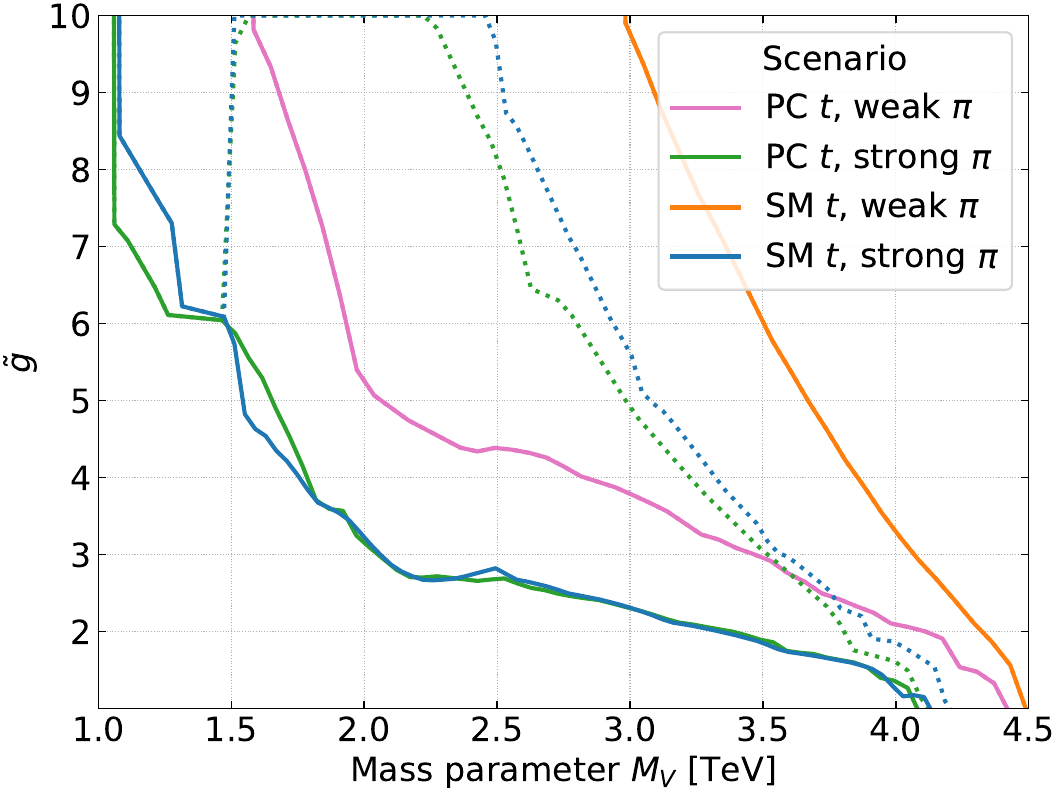}
        \caption{$\SU(5)/\SO(5)$}\label{fig:envelopes_su5so5}
    \end{subfigure}
    \begin{subfigure}{0.47\linewidth}
        \includegraphics[width=\linewidth]{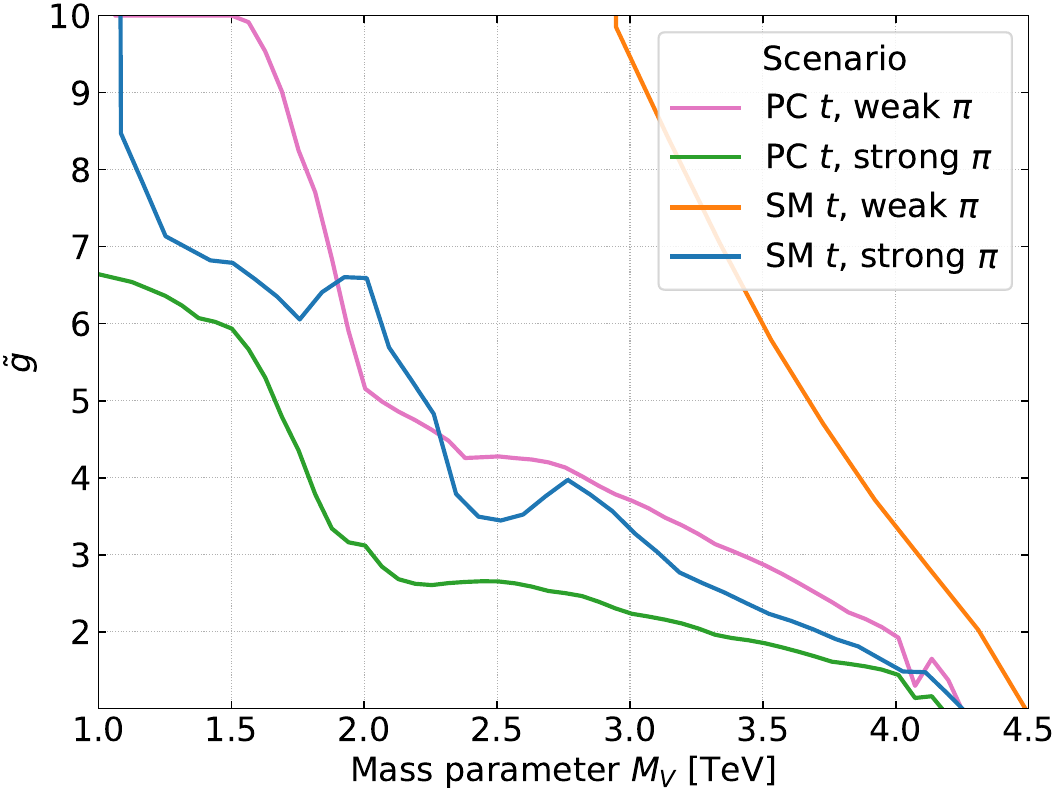}
        \caption{$\SU(4)/\Sp(4)$}\label{fig:envelopes_su4sp4}
    \end{subfigure}

    \caption{Left: bounds on the single production of heavy vectors in the $\SU(5)/\SO(5)$ coset. For 
    each scenario we show the envelope of the bounds from the individual channels shown in
    fig.~\ref{fig:allbounds_su5so5}, i.e.\ the strongest bound at every point. 
    The solid lines correspond to the fermiophilic, the dotted lines to the fermiophobic model, 
    both with $M_\pi = 700\,\mathrm{GeV}$.
    Right: corresponding bounds for the $\SU(4)/\Sp(4)$ coset for a pNGB masses of $M_\pi = 450$~GeV.   
    }
\end{figure}

The individual channels need to be combined to get the resulting bound on the underlying parameter 
space. We show in fig.~\ref{fig:envelopes_su5so5} the combined bounds for all four coupling scenarios
where each line represents the envelope of all channels. In the two scenarios with strong coupling to 
the pNGBs we show the fermiophilic scenario as a solid line and indicate where the fermiophobic case
differs by a dotted line. The scenario with weak couplings to top and pNGBs (orange line) is strongly
constrained yielding  $M_V > 3$~TeV -- 4.5~TeV depending on $\gt$. The bounds are considerably weaker, 
with $M_V$ down to 2~TeV remaining viable with only moderate $\tilde g$ provided only the PC 
couplings are turned on (pink line). The fermiophobic scenario is more strongly constrained than 
the fermiophilic one. The scenario with large  $\gV$ and SM like couplings to the top-quarks 
leaves the largest part of parameter space open, in particular in the fermiophilic case.
The corresponding results for the $\SU4)/\Sp(4)$ coset are shown in fig.~\ref{fig:envelopes_su4sp4}.
It differs only in the pNGB sector which features significantly less states as can be seen from
table~\ref{tab:cosets}. Again, in particular the scenario with no enhancement for the couplings to 
top quarks and pNGBs is strongly constrained.

\section{Conclusions and outlook} 
 
We have investigated the phenomenology of electroweak spin-1 resonances in
Composite Higgs Models paying particular attention to bounds from existing LHC data.  
The focus has been on models with gauge fermionic UV completions \cite{Ferretti:2016upr,Belyaev:2016ftv} 
because they provide detailed information on the quantum numbers and properties of the bound states.
More precisely, we have considered the cosets a $\SU(4)/\Sp(4)$ and $\SU(5)/\SO(5)$.
In view of the LHC phenomenology, the  states which can mix
with the electroweak vector bosons of the SM are the most constraint ones as they can be singly
produced. Independent of the coset there is always one charged spin-1 resonance mixing
sizably with the W-boson and two neutral spin-1 resonances mixing  sizably with the Z-boson. 
This is  a consequence of the fact that in all cases the unbroken subgroup
contains the custodial group $\SU(2)_L \times \SU(2)_R$ by construction. 

We have used various analyses to obtain bounds in the mass-coupling plane for these states.
Direct searches for heavy resonances in the s-channel at the LHC are used for decays of the
spin-1 resonances into two SM fermions. In addition we have used recast searches for final
states containing two bosons, either pNGBs and/or SM vector bosons as for these no direct searches had 
been performed. We have investigated four different cases to study the effect of unknown model 
dependent couplings. We have found, masses as low as 1.5~TeV are still allowed by current LHC data 
in scenarios with sizeable couplings of the spin-1 resonances to pNGBs.
In this part of the parameter space, also the states which only mix weakly or not at all will have 
masses in the same range. 

One might get further bounds or even discover these in processes such as
\begin{align}
  gg &\to b \bar{b} V^0, \,    t \bar{t} V^0 \\
    gg &\to b \bar{t} V^+,\,  t \bar{b} V^- \,.
\end{align}
Moreover, a projective future circular collider 
\cite{FCC:2025lpp,FCC:2025uan,FCC:2025jtd}
will offer  additional possibilities for the pair production of these states.

\section*{Acknowledgements}
\noindent
We thank the organizers of the workshop for provided such an stimulating environment.
This work has been supported by DFG, project no.~PO-1337/12-1 as well as the  research training group GRK-2994.

\end{document}